**Accessibility optimization of public transportation in historical districts – a study of Bei'lin District, Xi'an**

Shengxuan Ding[1] and Changwei Yuan[2 *] and Jing Chen[3] and Chenhao Zhang[4] and Jian Feng[5]

[1] School of transportation engineering, Chang'an University, Xi'an,710061; PH 184-0519-6669; E-Mail: 2020234075@chd.edu.cn

[2] (Corresponding Author) School of transportation engineering, Chang'an University, Xi'an, 710061; PH 139-9280-2916; E-Mail: changwei@chd.edu.cn

[3] School of transportation engineering, Chang'an University, Xi'an, 710061; PH 186-2937-3382; E-Mail: chenjing@chd.edu.cn

[4] School of transportation, Southeast University, Nanjing, 211189; PH 139-5171-6936; E-Mail: 1071955807@qq.com

[5] School of transportation engineering, Chang'an University, Xi'an, 710061; PH 136-3021-6091; E-Mail: 2021234097@chd.edu.cn

## ABSTRACT

With continuous improvement of urbanization and motorization, travel demand in historical blocks is higher than before. The contradiction between supply of transportation facilities and environmental protection is more serious. Traditional public transport planning methods aim to improve mobility. However, several existing studies do not put travelers at the first place, which ignore quantitative description of land uses and the characteristics of travelers. This paper intends to improve individual accessibility of public transportation in historical districts. Based on POI data, the calculation of accessibility combines the utility model and spatial interaction model. To optimize public transportation, the goal of improving individual accessibility is transformed into reducing residents' travel negative utility. Using a real world dataset in Bei'lin District of Xi'an, the performance of the proposed model is evaluated. This paper calculates the present accessibility and uses ant colony algorithm to optimize public transportation. The results demonstrate that the calculation and optimization of public transportation accessibility are practical, which are also valuable to public transportation planning and organization in historical blocks.

## INTRODUCTION

Historical districts integrate basic business, residents' life and entertainment. The demand of travel is higher and higher, and historical districts are usually located in the central area of the city. The narrow road network often becomes the bottleneck of traffic development, which leads to the contradiction between the supply of traffic facilities, environment protection and travel demand. With the advantages of large transportation volume, environmental protection and low carbon, public transportation is an effective way to solve the contradiction between travel demand and block protection. Therefore, on the premise of maintaining the pattern of historical blocks, optimizing the layout and accessibility of public transportation is an important way to ensure the orderly operation of transportation and the sustainable development of historical blocks.

This paper takes Bei'lin district in Xi'an as an example, and focuses on the current comprehensive traffic situation and accessibility of public transportation. Based on the analysis of the current traffic situation, this paper constructs a public transport accessibility measurement model and optimizes the existing routes, so as to improve the accessibility of public transportation. To improve the accessibility of public transportation, optimization can make individual travelers more convenient. At the same time, the results can provide more forward-looking suggestions for decision makers, which are in line with the sustainable transportation development strategy in historical districts.

## LITERATURE REVIEW

In terms of the measurement and optimization of public transport accessibility, Yang(2019) evaluated the public transport accessibility by considering the public transport operation, station configuration and surrounding environment. The comprehensive service frequency of each POI is calculated, and the indicators are converted into community level. Cao(2018) evaluated the accessibility of urban road network public transport by using two indicators, the supply level of urban public transport facilities resources and the travel time ratio of private bus. Walking time of pedestrians to the bus station and the spatial layout of the bus station are considered. Tao(2020) discussed the influence of public transport on working accessibility by multi-mode method. It is found that the accessibility of public transportation is much worse than that of cars and the distribution of public transportation is more uneven. The paper emphasizes countermeasures of land use and transportation policies to improve the accessibility of public transport. Jiang (2020) constructed model and analyzed four kinds of bus operation strategies from the perspective of travel time and accessibility. This paper compared the difference between travel time under different operation strategies by analyzing the model. When the travel demand is at low level, the responsive bus with independent lines equips with the highest accessibility. When the travel demand is at a high level, the fixed route bus with fixed stops equips with the highest accessibility. Xu(2018)  proposed an index to measure whether urban land use planning considers public transport accessibility, and measured the spatial accessibility of bus stops by the number of effective access grids. Using geographically weighted regression model, the correlation between traffic accessibility and urban land use characteristics was determined. Rui (2018) established a set of spatial accessibility evaluation method of multi-mode transportation system which takes Shanghai Hong'qiao traffic hub and its surrounding areas as an example. The total travel time of walking, waiting and transfer is taken as the index, which provides a simple and flexible method for the spatial evaluation. It also measures network transformation of public transport system accessibility under various modes of transportation.

In the aspect of traffic optimization in historical blocks, Jian, Zhuo (2017) proposed the concept, measurement and possible application of traffic calming in historic conservation district in China. According to the research of Qiuping Wang eta.(2020) , in order to solve the problems of unreasonable layout and planning in historic districts, she proposed a game model to figure out the best road space division plan, which minimized the generalized travel cost. in the mixed road section of a historical district. Qiuping Wang eta. (2019)  established a dynamic game logit model of traffic competition, which considering sharing rate of different traffic modes by maximizing the generalized profit of economy, environment, road service level. This paper verified the proposed method by the case of Academy Street in Zhengzhou city, China. What's more, Ou Zheng (2022) provides a framework to process trajectory data, which can be applied in analyzing public transportation by data processing. The Multi-Modal Large Language Model is not only applied in the area of public transportation, but also in the field of medicine (Huang et al., 2023).

To sum up, most of the existing studies about evaluation of public transport network accessibility are based on the relationship between land use and accessibility. Most of them aim at improve the efficiency of road network system or public transportation system, ignoring quantitative description of land use, attractiveness and traveler characteristics. In the aspect of optimization of public transportation accessibility, the qualitative optimization scheme targets at the public transport network, ignoring quantitative optimization. The characterization of travel demand of specific individuals in historical blocks needs further study. In the aspect of public transportation in historical districts, many researches focus on the analysis of the spatial proportion of different types of transportation modes, the characteristics of travelers and the hierarchical optimization of the road space.. It can be seen that researches on individual travel needs to be further studied and expanded. Therefore, in order to improve the convenience of residents , this paper measures and quantitatively optimizes the accessibility of public transportation in historical districts based on available POI data.

## Methodology

As stated in the literature, accessibility in historical districts can be described as the convenience of individual travelers to complete specific activities from the starting point to the destination by using urban public transportation within a specific time. This paper analyzes and evaluates the accessibility of public transportation in historic districts from the aspects of land use, transportation facilities, space-time variables and individual factors.

## Model composition

This paper combines the spatial and utility model which can analyze the travel positive and negative utility to measure the accessibility.

(1) Calculation of positive utility

Compared with OD matrix, POI index consists spatial coordinates, category, classification and other information, which can describe the spatial characteristics abundantly. The scale, type and density of distribution in historical district can be described by POI index well.

The positive effects of the model is calculated by POI index:

$$D_i = \sum_{e=1}^{E} \omega_e \times d_{ie} \tag{1}$$

Di —— Equivalent of poi in traffic district i;

E —— Categories of POI;

$\omega_e$ —— Weight of class POI e;

$d_{ie}$ —— Number of POI e in traffic district i

（2）Calculation of negative utility

Considering psychological characteristics and travel demand of travelers in historical districts, the parameters of travel time are calibrated according to the results of travel questionnaire survey, and appropriate weights are given to calculate the negative utility.

The negative utility is characterized by walking time, waiting time and traveling time as follows:

$$C_{ij} = \partial_1 \times T_1 + \partial_2 \times T_2 + \partial_3 \times T_3 \tag{2}$$

$T_1$ —— Walking time;

$T_2$ —— Waiting time;

$T_3$ —— Travel time onboard;

$\partial_i$ —— Weight of time spent in each stage.

The calculation methods of each index are as follows：

$$T_1 = \frac{L_O + L_D}{V} \tag{3}$$

$L_O$——Distance from starting point to a bus stop;

$L_D$——Distance from ending point to destination;

V——Walking speed, 1. 2m/s

$$T_2 = \sum_i^m \frac{(1/f_i)^2 + \delta^2}{2} \tag{4}$$

$f_i$——Frequency of buses, veh/h;

m——Number of trips, in this paper, due to zero transfer, the number of trips is taken as 1;

$\delta$——Deviation factor, deviation between actual departure time and schedule, 0. 707/fi.

It is assumed that the starting and ending stations of personal travel are the same bus stop, and the two stops are on the same bus line.

$$T_3 = \frac{\sum_{ij}^n \frac{L_{ij}}{v}}{n} \tag{5}$$

n——Number of bus routes between stations;

$L_{ij}$——Distance between stations;

v——Average speed between bus stops, 20km/h.

To sum up, the calculation of negative utility in public transportation is obtained as follows:

$$C_{ij} = \partial_1 \times \frac{L_O + L_D}{V} + \partial_2 \times \sum_i^m \frac{(1/f_i)^2 + \delta^2}{2} + \partial_3 \times \frac{\sum_{ij}^n \frac{L_{ij}}{v}}{n} \tag{6}$$

**Model properties and formulation**

This section aims at the optimization of the public transportation, which transforms the improvement of accessibility into the reduction of residents' travel negative utility.

$$min\, C_i = \sum_i^j C_{ij} \tag{7}$$

Referring to the max walking distance and the bus operation efficiency accepted by residents, the average operation speed of the bus line is set as 20km / h. From the aspects of line length, average station distance, line network density, line non-linear coefficient, average distance between stations, constraints are set as follows:

St:

$$5\ km \le L \le 20\ km \tag{8}$$

$$0.3\ km \le l \le 0.8\ km \tag{9}$$

$$P = \frac{L}{S} \ge 3\ km/km^2 \tag{10}$$

$$NL = \frac{L}{d} \le 1.4 \tag{11}$$

$$1.5 \le b \le 2.5 \tag{12}$$

$$T \le T_{max} \tag{13}$$

L——Length of bus line

l——Average distance between stations, $l = \frac{L}{\text{n-1}}$

n——Number of stops

P——Network density

S——Study area

d——Straight line distance between OD

b——Line repetition factor

**Solution approach**

According to the layout and operation characteristics of public transportation, in order to improve individual accessibility, this paper modifies the formation of ant colony algorithm code. Based on the principle and characteristics of ant colony algorithm, the solving steps of the optimization of public transportation line are proposed as follows:

Parameters setting of ant colony algorithm:

M —— Number of ants;

$n_{ij}$ —— The visibility between points i and j, expressed by the reciprocal of the objective function, which represents the expected value of the ant transferring from station i to station j;

$t_{ij}$——The pheromone strength of the line from i to j;

□ $t_{ij}^{k}$ —— Pheromone increment left by the ant k in the loop solution. If ant k chooses the path (i, j), then Δk = C1; if ant k does not choose the path (i, j), □ $t_{ij}^{k}$= 0. Among them, C1 uses constant to express pheromone strength;

$P_{ij}^{k}$ ——Probability that ant k selects a route in multiple sites, $Pij^k = \frac{tij^{\partial} \times nij^{\beta}}{\sum tij^{\partial} \times nij^{\beta}}$. j is the optional node that ant k faces; α is used to describe the importance of trajectory; β is used to describe the importance of visibility.

(1) Transfer rules

In the process of optimization, ants tend to move towards the direction with the highest pheromone concentration. According to the path left by ant colony and the initial information of pheromone, they choose the direction with higher probability.

(2) Pheromone update

The pheromone left by ants will volatilize. Therefore, when calculating pheromone accumulation, after all nodes are visited by ants, the residual pheromone is updated, and the vocalization rate of pheromone trajectory is introduced to adjust the amount of information on each path within time t as follows:

$$Tij(n+1) = (1-p) \times tij(n) + \sum \Delta\, tij(n) \tag{14}$$

Optimization steps:

The core of ant colony algorithm is to select pheromone of ant. In this paper, the negative utility of travelers is considered as the main factor of the objective function for optimization:

$$Ci = {t_{ij}}^{\partial} \times {n_{ij}}^{\beta} \tag{15}$$

where, Ci is the pheromone left by ants in the process of optimization;

${n_{ij}}^{\beta}$ is defined as the reciprocal of time, α indicates the importance of the trajectory; β indicates the importance of visibility.

## CASE STUDY

### Measurement model

Taking Bei'lin district of Xi'an as an example, this paper divides it into 16 traffic districts according to land use properties and attractiveness, which is shown in Figure 1.With the help of ArcGIS platform, the center position of the traffic district is obtained by the center of gravity method as the starting point of a travel. The station nearest to the district center is obtained by the nearest neighbor analysis method in ArcGIS, which is used as the starting point onboard.

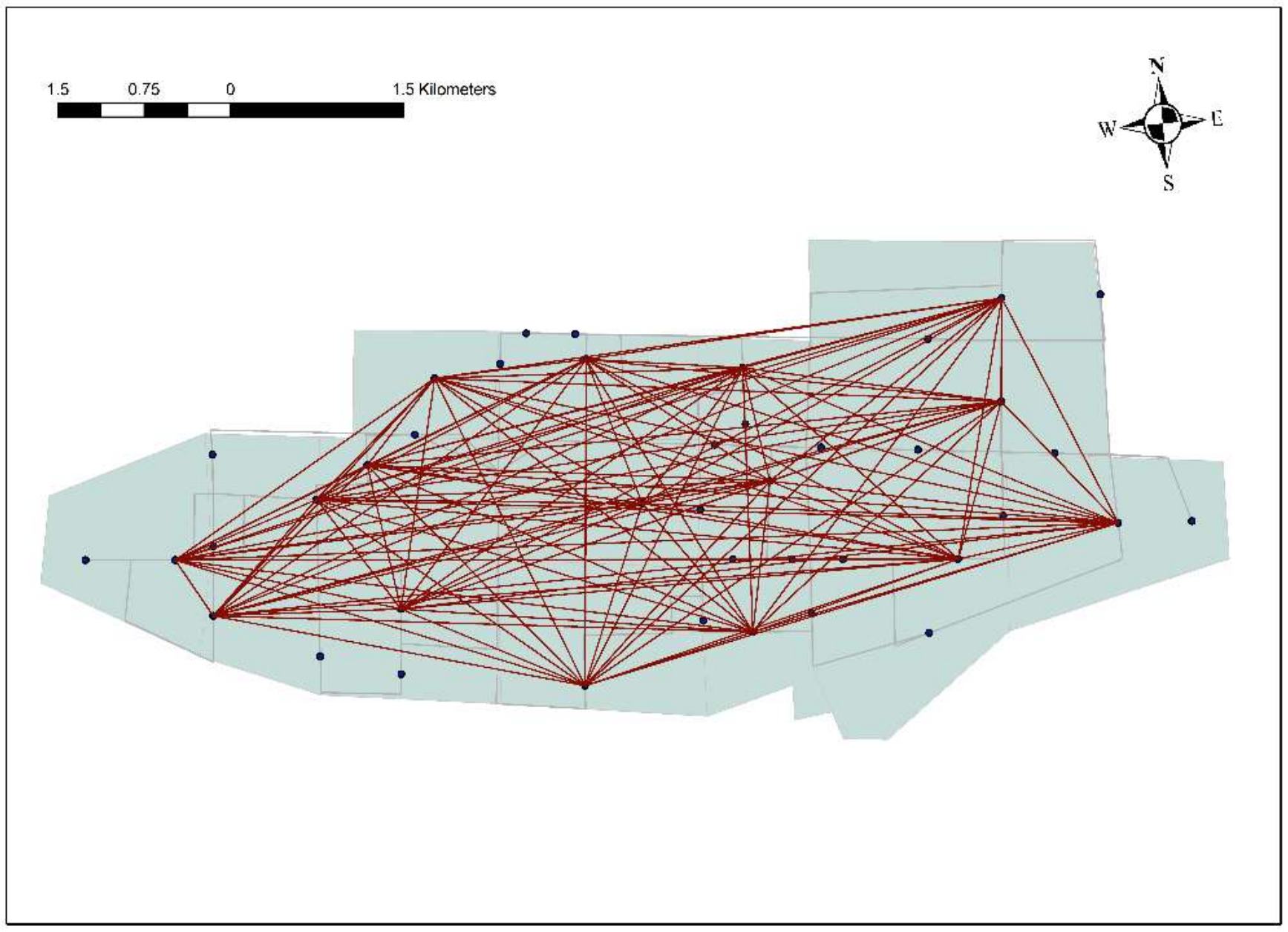

**Figure 1. Traffic district division and OD distribution**

In this paper, the topological relationship between public transportation network and bus stops is constructed. After verification and correction, public transportation data set is established, and the OD distance cost matrix between traffic districts is generated. After verification, the number of starting points multiplied by the number of ending points is equal to the number of lines. Then, using the path analysis in ArcGIS, the length of bus travel path between stations and travel time are calculated, which is shown in Figure 2.

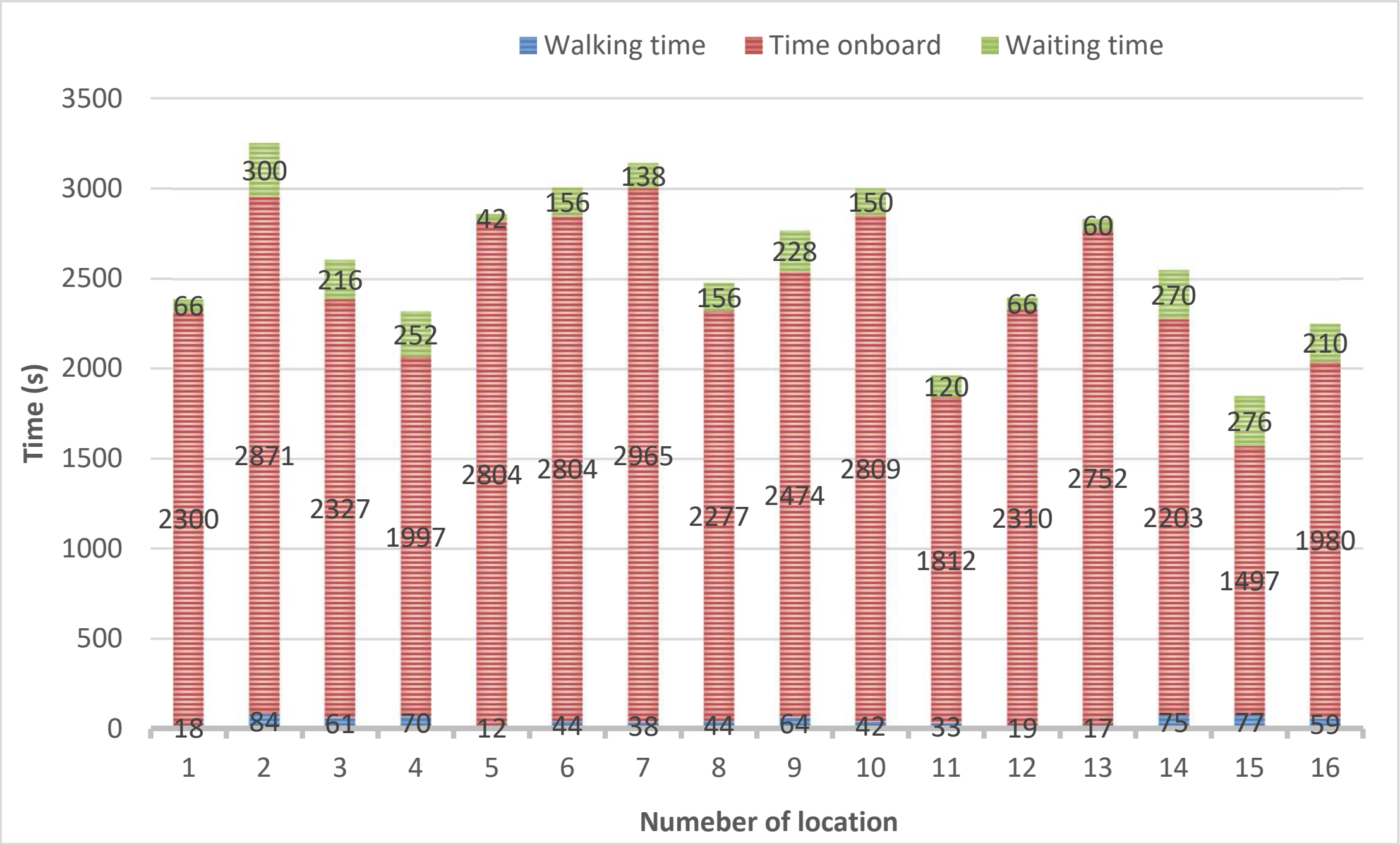

**Figure 2. Weighted travel time of each stage**

The data of interest points in this paper is oriented from Gaode map in 2019. After merging and classifying the POI data, a total of 13 categories and 44057 items are selected, the weight and quantity of POI data are shown in Table 1. According to POI index, the positive utility of each traffic district is calculated.

**Table 1 Weight and quantity of POI**

| NO. | Type | Weight | Quantity |
|---|---|---|---|
| 1 | Food and Beverages | 0. 0392 | 9689 |
| 2 | Scenery | 0. 1079 | 163 |
| 3 | Public facilities | 0. 0361 | 378 |
| 4 | Shopping service | 0. 0722 | 12618 |
| 5 | Business residence | 0. 0560 | 1599 |
| 6 | Life service | 0. 1150 | 8248 |
| 7 | Financial and insurance services | 0. 0988 | 995 |
| 8 | Science, education and cultural services | 0. 0912 | 2935 |
| 9 | Sports leisure service | 0. 0590 | 1111 |
| 10 | Medical service | 0. 1128 | 1405 |
| 11 | Government agencies and social organizations | 0. 0819 | 1119 |
| 12 | Accommodation services | 0. 0939 | 1608 |
| 13 | Transportation facilities | 0. 0361 | 2189 |

Combined with the qualitative analysis of actual land use properties, we can figure out that the current accessibility calculated by the evaluation model is consistent with current situation. In conclusion, the accessibility of traffic District 5, 7 and 15 are the lowest in study area, which is the main object to optimize in this paper.

## Optimization based on ant colony algorithm

Considering the current layout of public transportation, the optimized bus routes are determined as follows in Figure 3:

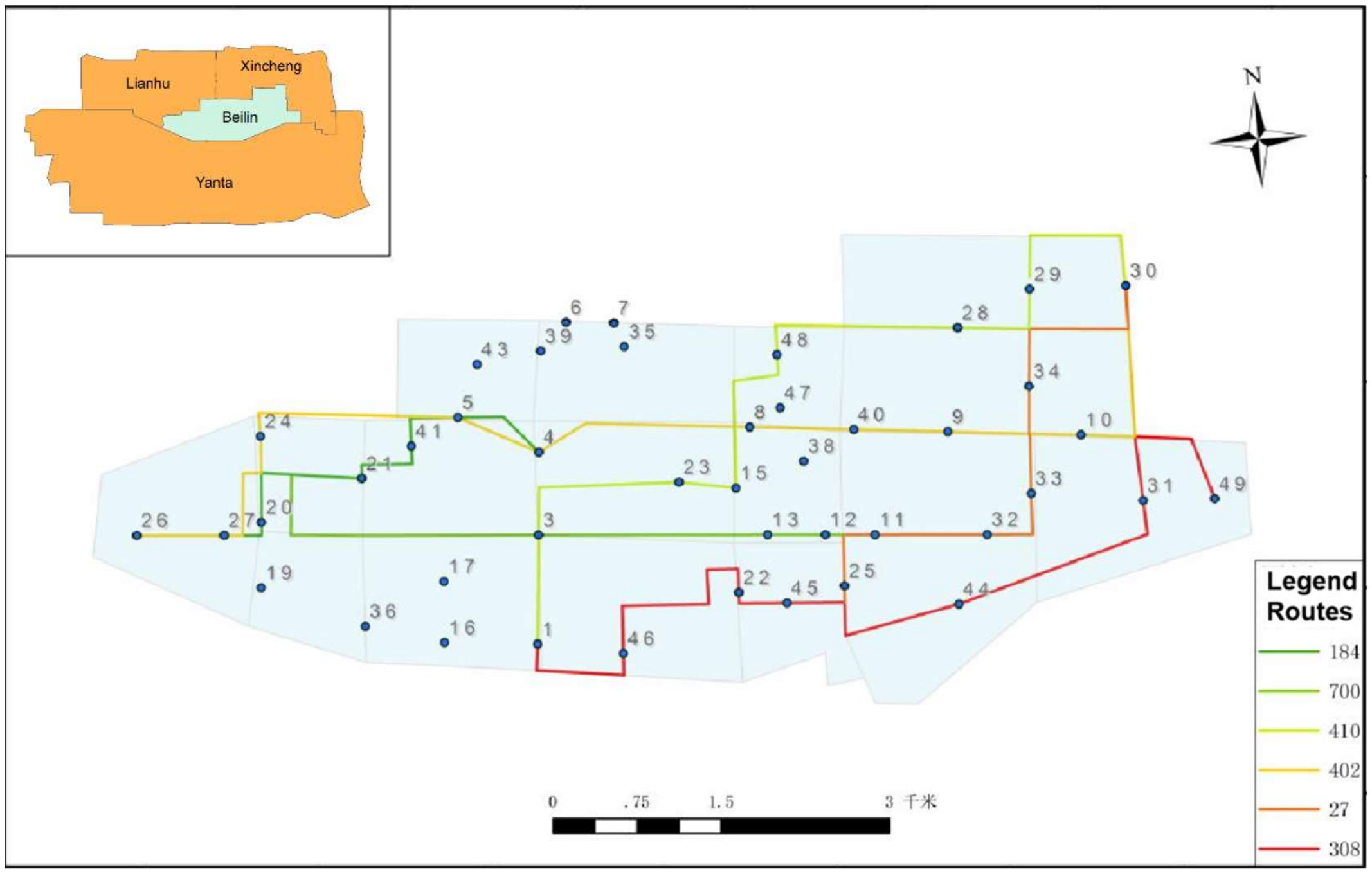

**Figure 3. Starting and ending points and bus lines need to be optimized**

With the help of MATLAB ,this paper selects ant colony algorithm to optimize the layout of bus routes, the optimal solution of the model is obtained through iteration. After verifying whether the layout line meets the constraint conditions, the layout scheme of public transportation is shown in Table 2 and Figure 4:

**Table 2 Starting and ending points after optimization**

| ID | Station | Length（km） |
|---|---|---|
| 26-29 | 26-27-21-41-5-43-39-35-48-47-40-9-34-29 | 13. 97 |
| 26-32 | 26-27-21-41-3-5-46-45-25-11-32-31 | 10. 62 |
| 46-30 | 46-45-25-11-32-31-10-34-30-30 | 12. 11 |

**Result analysis**

After calculation and verification, the optimized line length, average station distance, line network density, line non-linear coefficient and other indicators meet the constraints. It can be seen that the optimized indicators meet the constraints and are better than the current situation, the result is shown in Figure3. and Figure 4.

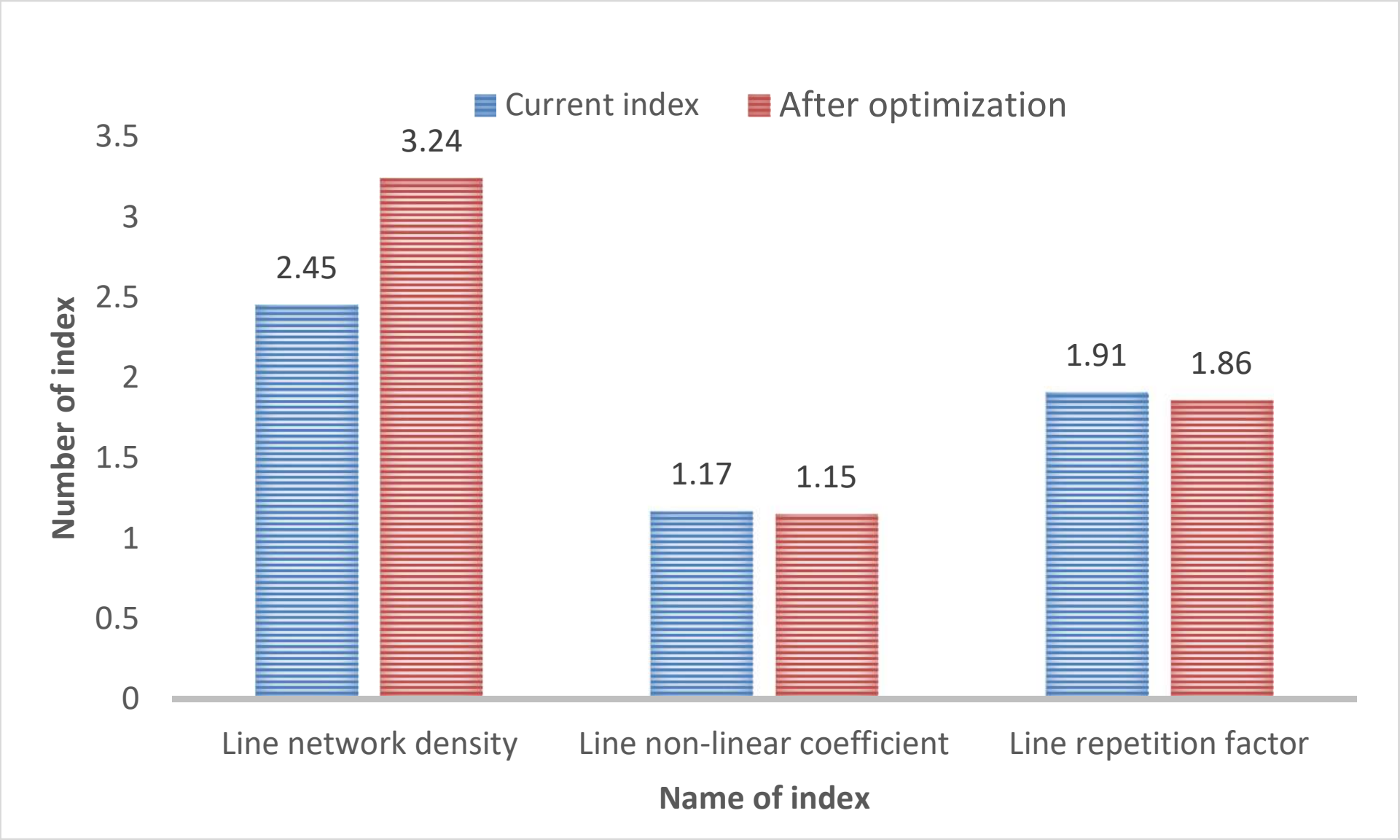

**Figure 4. Feature comparison before and after optimization**

Through calculation of the accessibility of public transportation after optimization, the accessibility has been significantly improved, especially traffic districts at low level of accessibility. The calculation is shown in Figure 5:

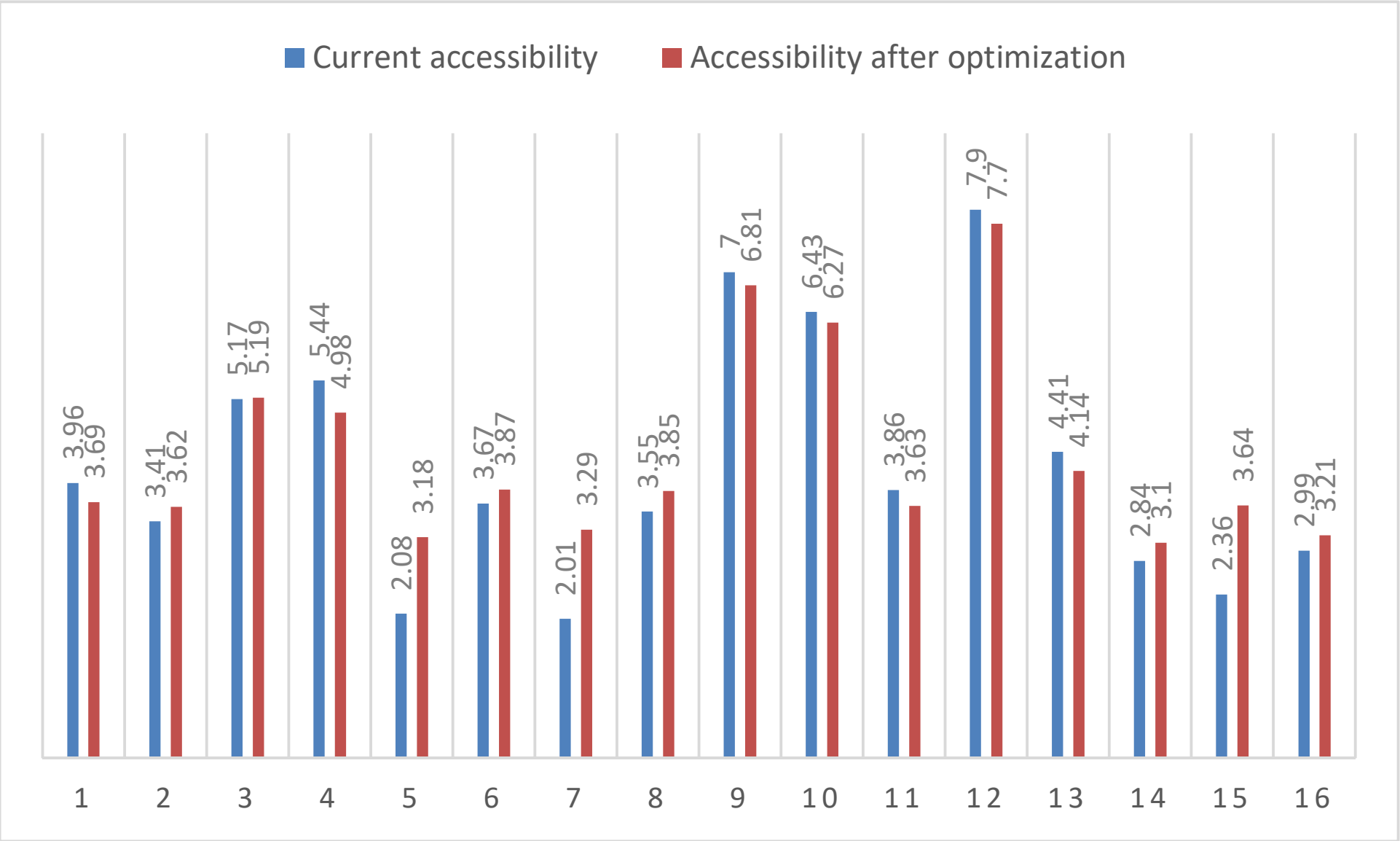

**Figure 5. Accessibility before and after optimization**

## Conclusion and future work

Taking Bei'lin District of Xi'an as an example, this paper analyzes and optimizes the comprehensive traffic status and the accessibility of public transportation in historical district of ancient city. This paper takes unique land properties in historical districts into account. Due to the advantages of POI data, which is more accessible than OD data to characterize the accessibility, this paper selects POI to measure accessibility and uses ant colony algorithm to quantitatively optimize the public transportation, considering the traveler characteristics. The connection and transfer of public transportation should be considered in future studies to improve the performance of calculation and optimization. In addition, more practical variables to measure accessibility can be selected. The environmental protection of historical districts can be combined with the layout of public transportation.

## Acknowledgment

This research was supported by the National Key Research and Development Program of China (No. 2020YFC1512004) and the Foundation for Distinguished Young Scholars of Shaanxi (No. 310823160103).